Title: Bond polarons and high-Tc superconductivity in single layer La$_{2-x}$Sr$_x$CuO$_4$: normal state currents and pairing.
Author: Mladen Georgiev {Institute of Solid State Physics, Bulgarian Academy of Sciences. 1784 Sofia, Bulgaria}
Comments: 17 pages with 3 figures and 1 table, all pdf format
Subj-class: physics

We use the term *bond polaron* to designate the phonon-coupled entity which materializes the link between neighboring conductive CuO$_2$ layers in high-Tc superconductive materials. This link is essential for the superconductivity which requires long range phase coherence in addition to pairing of carriers. The linkage features have been studied in detail and the results will be discussed below as will the implications for the superconductive mechanism. These point to a process which is less dependent on the doped hole density implanted as *x* from the solid solution. The present results apply to single layer materials mostly though some of them may be helpful to understanding sister multilayer materials as well.

1. Introduction

There are two main pre-requisites of a high-Tc superconductivity in oxocuprates: pairing and long-range phase coherence. There being many conceivable pairing models, we will confine our discussion to the ones closest to our line of reasoning throughout this paper: the Van-der-Waals or polarizability-bound models. The vibronic polarizability arises due to the occurrence of inversion dipoles because of the violation of inversion symmetry at the vibronic sites. Once there, the vibronic polarizability gives rise to an s-atom like structure which is electrostatically neutral even though highly polarizable. These atomic-like species can bind forming molecules much like s-atoms do[1-4].

In the solid state language, single polarizable vibronic species should be called small polarons if they produce deformations on an atomic scale just like any Jahn-Teller polaron. These species can bind in momentum space to form vibronic Cooper pairs or in real space to form vibronic bipolarons. In so far as these polarons travel along O-O bonds, we call them *bond polarons* or *Cooper's (semi-) pairs*.

The other high-Tc superconductivity component, the long-range phase coherence, requires a sufficiently effective linkage between conductive CuO$_2$ planes. Such is provided by the hole transfer from O$_{planar}$ to O$_{apex}$ site through carrier travel in La$_{2-x}$Sr$_x$CuO$_4$. This transfer may be purely electronic in character but it proves to be phonon coupled, as seen shortly. Carrier transport along a c-axis bond may couple to A$_{2u}$ or E$_u$ modes to become phonon coupled. The distortion cloud along or normal to a bond should be similar to general deformation shapes.

The temperature dependences ρ(*T*) of the electric resistivities $\rho_c$ (axial) and $\rho_{ab}$ (in plane) in normal state of single-layer high-$T_c$ superconductors, such as the $La_{1-x}Sr_xCuO_4$ and $Bi_2Sr_{2-x}La_xCuO_y$ families, has been measured to show common trends: As the temperature *T* is raised, the resistivity first drops steeply (*dρ/dT*<0) before it starts rising (*dρ/dT*>0) as ρ ∝ *T* above an apparent semiconductor-to-metal crossover temperature. To analyze ρ(*T*), we plotted *T* / ρ against *T* at various doping levels *x* for both $\rho_c$ and $\rho_{ab}$. In ionic theory, T/ρ is proportional to the average traversal time across a potential energy barrier as an ionic particle undergoes a drift in an external electric field [5-7].

We've found T/ρ to be in good agreement with the temperature dependence of the quantum mechanical rate of a migrating ionic particle: It exhibits a zero-point rate at the lowest temperatures, followed, as T is raised, by a relatively insensitive flat plateau before a transitory range leads to the Arrhenius temperature activated branch. We have found evidence for the admixture of 0-phonon, 1-phonon, and eventually 2-phonon absorption processes at temperatures before the Arrhenius range. These features shape the semiconductor branch below the semiconductor-to-metal crossover. As the semiconductor-like branch has become less efficient on raising the temperature, the metallic-like branch has taken over predominating, its ∝ *T* character deriving from the coupling of the migrating particle to the external electric field. The general concord indicates that traditional metal physics may not suffice, as long as ionic features play an essential role in transport phenomena. We attribute our conclusions to the drift of strong-coupling polarons moving along either axial or in-plane Cu-O bonds. These *bond polarons* have originated from carrier scattering by double-well potentials associated with the Cu-O bonds. A bond polaron has dissociated into a free hole as it has been leaving the Cu-O bond to step over a neighboring O-O link [8-11].

Herein we readdress data on the temperature and doping dependencies of normal state currents in single layer superconductors of the $La_{2-x}Sr_xCuO_4$ family. We follow previous lines of insight into energy gap, coupled frequency and barrier to complement earlier analyses [12-19].

2. Axial (in-plane) resistivity.

For analyzing the resistivity, we followed a way based on the classical definition of a temperature-dependent ionic "*drift velocity*". See the unit cell of $La_{2-x}Sr_xCuO_4$ for an illustration of the conceivable steps of a charge leak along the c-axis: There are two essentially different steps. Accordingly, the axial electrical resistivity is obtained from:

$$\rho_c = 1/\sigma_I + 1/4\sigma_{II} + 1/4\sigma_{III}, \tag{1}$$

while $\sigma_I$ and $\sigma_{III}$ are defined as:(cf. [7])

$$\sigma_I = \sigma_{III} = Ne\, R_{Cu(P)\text{-}O(A)}\, k_{12}(T)\{\exp(\boldsymbol{p}.\boldsymbol{F}/2k_BT) - \exp(-\boldsymbol{p}.\boldsymbol{F}/2k_BT)\}/F$$

$$\approx Ne\,[R_{Cu(P)\text{-}O(A)}\,k_{12}(T)(p/k_BT)] \quad \text{at } \boldsymbol{p}.\boldsymbol{F} = pF \ll 2k_BT \tag{2}$$

where $\boldsymbol{F}$ is the external electric field applied along the c-axis. We also proposed the following approach to $\sigma_{II}$:

$$\sigma_{II} = Ne\, R_{O(A1)\text{-}O(A3)}(t_{pp}/h)\cos\theta. \tag{3}$$

$N$ is the axial carrier density and $e$ is the electronic charge, $R_{Cu\text{-}O}$ and $R_{O\text{-}O}$ are the respective bond lengths, $t_{pd}$ and $t_{pp}$ are hole hopping terms. Using Coulomb potentials, apical oxygen $2p_{x,y,z}$ orbitals, and copper $3d z^2$ orbitals to calculate the hopping energies, we show that $t_{pp} \gg t_{pd}$ which implies that the Cu(P)-O(A) step is the bottleneck of the axial transport. $p$ is an electric dipole associated with the Cu(P)-O(A) CT.[20] The temperature dependence of $\rho_c$ is basically carried by $k_{12}(T)$ which is a dynamically-coupled barrier-controlled CT rate described at length elsewhere.[7,21,22] The expression within square brackets in $\sigma_I \equiv \sigma_{III}$ is the apparent carrier mobility $\mu_I$. In an ionic crystal approach appropriate for events along the c-axis, the mobility is characteristic of a coherent polaronic transport towards the apex sites:

Accordingly, we define the drift velocity of a carrier scattered by the double-well barrier as the ratio of the respective bond length to the average barrier-traversal time $\tau_{12}(\boldsymbol{F}) = [2k_{12}(T)\sinh(\boldsymbol{p}.\boldsymbol{F}/2k_BT)]^{-1}$. $R_{Cu(P)\text{-}O(A)}$ is the Cu(P)-O(A) bond length, $R_{O(A1)\text{-}O(A3)}$ is the O(A)-O(A) bond length of two apex oxygens across a rocksalt layer, $\theta$ is the angle which the $O(A_1)\text{-}O(A_3)$ inter-connecting line concludes with the c-axis.

Undoubtedly, the agreement of our theoretical rates with the experimental $T/\rho(T)$ data gives evidence for the formation of *bond polarons* during the bottleneck steps of the axial charge transfer. Our bond polaron transfers charge from a Cu(P) site to an adjacent O(A) site giving rise to the observed current. At the end, the bond polaron decomposes to a *free hole* at the O(A)-O(A) link step before reappearing again in a subsequent $CuO_6$ octahedron.

3. Phonon coupled rate

3.1. Axial (in-plane) Hamiltonian

For introducing the rate $k_{12} = \tau_{12}^{-1}$, it will be feasible to consider a two-site Hamiltonian describing the physics of a phonon-coupled electron transfer between two orbitals:[23]

$$H_{PJT} = \sum_{(ij)\alpha\beta} t_{ij\alpha\beta}(Q_i,Q_j) a_{i\alpha}^\dagger a_{j\beta} + \sum_{n\alpha} \varepsilon_{n\alpha} a_{n\alpha}^\dagger a_{n\alpha} +$$

$$\sum_{n\alpha\beta} G_{n\alpha\beta} Q_n (a_{n\alpha}^\dagger a_{n\beta} + \text{c.c.}) + S\sum_n (\boldsymbol{P}_n^2/M_n + K_n Q_n^2) \qquad (4)$$

where i,j,n are site labels, $\alpha$ and $\beta$ = 1,2 are labels of the electronic energy levels. $a_{n\alpha}$, etc. are fermion annihilation operators in site representation, $t_{ij\alpha\beta}$ and $\varepsilon_{n\alpha}$ are fermion hopping and local energy terms, respectively. $G_{n\alpha\beta}$ are the fermion-phonon mixing constants, $\boldsymbol{P}_n$, $K_n = M_n \omega_{bare}^2$, $M_n$, and $\omega_{bare}$ are correspondingly the momentum, stiffness, reduced mass, and bare vibrational frequency of the coupled phonon mode with coordinates $Q_n$, etc. Although a single mode is assumed for simplicity, multimode generalizations are straight-forward. Allowances are made for both local and nonlocal electron-mode coupling via the dependences on the mode coordinates of the electron-mixing part and the electron hopping terms, respectively.

The complete Hamiltonian should read $H = H_{PJT} + H_{ab}$ where $H_{ab}$ incorporates hopping and local energy terms built up by in-plane operators, such as $a_{n\gamma}$ ($\gamma = 3d_{x^2-y^2}$, $3d_{z^2}$, $2p_{x,y}$, etc.), as well as a hopping term mixing in-plane $3d_{x^2-y^2}$ ($b_{1g}$ symmetry) holes with $3d_{z^2}$ ($a_{1g}$ symmetry) holes. The tetragonal symmetry notation in brackets reminds that not only copper orbitals but also oxygen ligands of the proper symmetry may be involved. Solving for the eigenvalues of $H$ would give energy bands, as found elsewhere.[24] However under the conditions of an axial hole hopping much slower than its in-plane counterpart, the axial component $H_{PJT}$ can be dealt with separately. Alternatively, in view of the low occupancy of the Cu(P) $3d_{z^2}$ level, hole hopping from the planar orbitals to $3d_{z^2}$ is but a small perturbation and the main axial transport features may be obtained to the zeroth-order approximation by solving for $H_{PJT}$ alone. We will postpone for a further study solving for the higher-order eigenstates. In any event, $H_{PJT}$ will account for the main features of the c-axis transport unless $3d_{z^2}$ holes mix too much with the $3d_{x^2-y^2}$ band holes which does not appear to be the case. Alternative axial leaks are conceivable via the direct coupling of in-plane $a_{1g}$ or $b_{1g}$ band holes to $2p_z$ ($a_{1u}$ symmetry) O(A) orbitals.

3.2. Axial (in-plane) rate

The two-site rate of horizontal energy conserving transitions in a strongly-quantized vibronic system at sufficiently small η reads:[21]

$$k_{12}(T) = 2\nu\sinh(\hbar\omega/2k_BT)\{\sum_{E(n)\gg\varepsilon(B)}\{2[1-\exp(-2\pi\gamma_n)]/[2-\exp(-2\pi\gamma_n)]\}W_C\exp(-E_n/k_BT) +$$

$$\sum_{E(n)\ll\varepsilon(B)}\{2\pi\gamma_n^{2\gamma(n)-1}\exp(-2\gamma_n)/[\Gamma(\gamma_n)]^2\}\{\pi[F_{nn}(q_0,q_C)/2^n n!]^2 \times$$

$$\exp(-\varepsilon_R/\hbar\omega)\}\exp(-E_n/k_BT)\} \qquad (5)$$

where the expressions within small curled brackets are electron-transfer and configurational-tunneling probabilities, overbarrier for $E_n > \varepsilon_B$ and underbarrier for $E_n < \varepsilon_B$, respectively,

$$\gamma_n = (\varepsilon_{\alpha\beta}^2/8h\omega)(\varepsilon_R|E_n-\varepsilon_C|)^{-1/2} \tag{6}$$

is Landau-Zener's parameter,

$$F_{nn}(q_0,q_C) = 2q_0 H_n(q_C)H_n(q_C-2q_0) - 2nH_{n-1}(q_C)H_{n-1}(q_C-2q_0) +$$

$$2nH_n(q_C)H_{n-1}(q_C-2q_0) \tag{7}$$

is a quadratic form of Hermite polynomials, $\omega=2\pi\nu$ is the angular vibrational frequency. We set $W_C=1$ for the overbarrier tunneling probability. $q_0$ and $q_C$ are the scaled absolute well-bottom and crossover coordinates, respectively, $q = (K/h\omega)^{1/2}Q$, $Q$ being the actual configurational coordinate.

The remaining parameters controlling the isothermal rate are: $\varepsilon_R$, the lattice-relaxation energy, $\varepsilon_C$, the crossover energy, and $\varepsilon_B$, the interwell energy barrier, as in depicting a symmetric double-well potential. They relate to $\eta = \varepsilon_{\alpha\beta}/4\varepsilon_{JT}$ with $\varepsilon_{JT}=G^2/2K$ by way of:

$$\varepsilon_{JT} = \varepsilon_B / (1-\eta)^2$$

$$\varepsilon_{NB-AB} = 4\varepsilon_{JT}\eta$$

$$\varepsilon_R = 4\varepsilon_{JT}(1-\eta^2) = 4\varepsilon_B(1+\eta)/(1-\eta) = 4\varepsilon_C(1-\eta^2)/(1+\eta^2)$$

$$\varepsilon_C = \varepsilon_{JT}(1+\eta^2) = \varepsilon_B(1+\eta^2)/(1-\eta)^2$$

$$h\omega_{bare} = \eta\omega / \sqrt{(1-\eta^2)} \tag{8}$$

The dynamic CT rate $k_{pd}(T)$ depends on three fitting parameters; we presently took them to be $\eta$, $\varepsilon_B$, and $h\omega$.

The isothermal symmetry of the double well follows from the requirement that the current be reversible forth and back along the c-axis. Under the original $t_{dp} \ll t_{pp}$ condition the interwell barrier $\varepsilon_B$ can be regarded as an ionization barrier for the Cu(P)-O(A) bond.

4. Analyzing experimental resistivity data

4.1. Experimental layout

Extensive experimental studies of in-plane and out-of-plane resistivities in single crystals of $La_{2-x}Sr_xCuO_4$ have been reported lately as have some studies of $Bi_2Sr_{2-x}La_xCuO_y$.[13] The normal state range has been extended down to very low temperatures (below $T_c$) in a high external magnetic field to suppress the superconducting pair formation. Both $\rho_{ab}$ and $\rho_c$ have displayed similar temperature dependences at various doping levels: In the underdoped range, a descending (semiconductor-like) temperature dependence below a crossover minimum at several tens of K is followed by an ascending (metallic-like) $\propto T$ branch above the crossover. Curiously, the crossover temperature falls near the onset of the orthorhombic-to-tetragonal low-temperature structural phase transition in doped specimens. The descending branch is suppressed as $x$ is increased and nearly disappears at optimum doping. Consequently, the temperature behavior of both $\rho_{ab}$ and $\rho_c$ turns allout metallic, still very nearly following the $\propto T$ law in the overdoped range. The difference between $\rho_{ab}$ and $\rho_c$ is quantitative but the higher-temperature metallic branch is missing in $\rho_c$ at low $x$, there being a nearly temperature-independent resistivity range instead.

4.2. Identifying the fitting parameters

We interpret the experimental dependences in terms of equations (1)-(3) where the CT rate $k_{12}(T)$ is given by the reaction-rate formula (7), while the doping dependencies are attributed to a free-hole suppressed JT-elongation of the Cu(P)-O(A) bond length, as suggested elsewhere.[7] We note that both the insulating and metallic branches are accounted for by the dynamic-rate formula, as the former descending-with-$T$ branch results from the thermally-activated portion $\propto \exp(-\varepsilon_A/k_BT)$ on $k_{12}(T)$, while the latter ascending-with-$T$ branch originates from the $\propto T^{-1}$ field-coupling term. The field term is dominant when the exponential saturates for $k_BT \gg \varepsilon_A$. A crossover occurs at $T_{cross}$ obtained from $\nu_{eff}\exp(-\varepsilon_A/k_BT_{cross}) \approx p/2k_BT_{cross}$.

Our analysis relies on three fitting parameters: the barrier $\varepsilon_B$, the vibrational frequency $\nu$, and the reduced gap parameter $\eta$, as well as on the scaling factor $P_{fit}$. Of these, $\varepsilon_B$ was determined directly from the slope of the highest-temperature Arrhenius-branch rate. Apparently, the barrier was related to the draw-back action of the electron originating at the starting point on the going away hole. Estimating $\nu$ was perhaps less straightforward, though it was derived from $h\nu \sim 4k_BT_\nu$ where $T_\nu$ was at the start of the transition range connecting the low temperature plateau region (transitions from the lowest level at n = 0) with the thermally activated range (overbarrier transitions at large n). $4T_\nu$ was taken to be the temperature of complete occupation of the next higher vibrational level at n = 1 in a harmonic oscillator based analysis. Finally, the gap parameter $\eta$ of a nonadiabatic process was determined as $\sqrt{(P_{fit}/P_{th})}$, as explained below. All the three main parameters were subject to more or less substantial variations until the resulting fit was considered sufficiently good for the purpose.

We first checked an independent least-squares fit to a series of Boltzmann exponentials, namely $k_{12}(T)_{exp} \equiv T/\rho = \sum_{n=0 \div 5} A_n \exp(-n\xi/\tau_n)$ with $\xi = 1/T$ and $\tau_n = k_B/\hbar\omega_n$, allowing for anharmonic differences in the mode frequencies at the different energy levels as quantized by n. The fitting results for $\tau_n$ were seen fluctuating which stimulated us to further use the harmonic single-frequency fitting formula at $\omega_n = \omega$ instead. The latter formula was seen to do nearly as well, except for a slight mismatch in the transition region between the low-temperature and Arrhenius branches. Once the fitting parameters $A_n$ and $\tau_n$ (*viz.* $\hbar\omega_n$) have been obtained, we took $\varepsilon_B$ from the Arrhenius slopes and finally $\eta$ and $P_{fit}$ by comparing a series of theoretical rates $k_{12}(T)_{theor}$ at the values of $\hbar\omega_n$, $\varepsilon_B$ found but at variable $\eta$, $P_{fit}$ until one of these rates fitted close enough to the "experimental rate" $k_{12}(T)_{exp} \equiv T/\rho(T)$. This procedure gave fitting values for $\eta$ and $P_{fit}$.

The graphical results are shown where the experimental data on both $\rho_c$ and $\rho_{ab}$ represented in terms of their derived transfer rates are plotted against the temperature $T$ at various doping levels $x$ .[13] The solid lines are our 'best fits' of the reaction rate equation (7) to the experimental data giving

$$k_{12}(T) = P_{fit}[T/\rho(T)] \tag{9}$$

where $P_{fit}$ is a scaling factor. The fitting parameters are: $\varepsilon_B$, $\eta$, and h$\nu$. A reasonable adherence is manifested of the theoretical equation (7) to the experimental rate plots (11) in that the latter plots shaped the temperature dependence of a phonon-coupled dynamic rate exhibiting both a low-temperature tunneling portion and a higher-temperature thermally-activated Arrhenius branch.[8] The agreement with the rate formula within the entire temperature range lends support to our interpretation of the resistivity data.

---

We find no evidence for a logarithmic singularity at the lowest temperatures, as claimed elsewhere.[13] Instead, a peak occurs in the predicted $\rho(T)$ dependence, as the low temperature flat term is overweighed by the exponential terms at the onset of the thermally-activated portion. However, this peak is hard to confirm experimentally in so far as it appears near the onset of a superconductivity which withstands the pair-inhibiting magnetic field at low temperature. Yet, traces of it may be spotted in the observed $\rho_c(T)$ curves at the larger *x*.

In most of the observed rate dependences, both axial and in-plane ones, the low-temperature tunneling rate within the plateau range was found to increase slightly though linearly with the temperature. Instead, the theoretical rate of an isothermal process, such as the reversible charge transfer, is predicted to be very nearly temperature-independent over the entire low-temperature plateau range. Now, it is safe to attribute the linear rate growth to a 1-phonon assisted tunneling process shown to increase $\propto$ T. Indeed, from the the condition that the 1-phonon emission and absorption rates be proportional to the respective number of phonons

$\Delta n_{\Delta E} = [\exp(\Delta E/kT) - 1]^{-1}$: $k_{em} \propto \Delta E\,(\Delta n_{\Delta E} + 1)$, $k_{abs} \propto \Delta E\,\Delta n_{\Delta E}$, we get both rates $\propto T$ for $kT \gg \Delta E$. The relevant resistivity is proportional to the phonon-assisted tunneling rate at two-level systems, as in glasses:[27]

$$\tau_{PA}^{-1} = C\coth(\Delta/k_B T) \qquad (10)$$

where $C$ is a constant and $\Delta$ is the vibronic tunneling splitting. The formula reduces to $\tau_{PA}^{-1} = C(k_B/\Delta)T$ at $k_B T \gg \Delta$.

The vertical tunneling process being vanishing at zero point, we checked a correction to the isothermal rate of the form $\Delta k_{12}(T) = AT$ so that the overall low-temperature rate branch became $\sim AT + k_{12}(0)_{iso}$. The fitting values of the linear rate coefficients A and the zero-point isothermal rates $k_{12}(0)_{iso}$ are both listed in Table I, while fits obtained by means of the corrected rate formula

$$k_{12}(T)_{corr} = k_{12}(T)_{iso} + AT \qquad (11)$$

are shown. A more rigorous approach to multiphonon absorption and emission rates will be discussed in greater detail elsewhere. In Fig.4 we also see that deviations of the experimental rates from the latter formula occur and become more pronounced as the doping x is increased. Indeed, after being completely absent at low x, misfits appear near optimal doping in that the experimental data are increasingly underestimated by the theoretical rate. This implies the inclusion of another process not accounted for so far by the theoretical rate formula. The most obvious candidate is the 2-phonon vertical tunneling and, indeed, preliminary estimates showed that the discrepancies were proportional to higher powers of T.

At doping levels higher than optimal ($x \geq 0.15$) the critical temperature drops, $La_{2-x}Sr_xCuO_4$ becoming virtually nonsuperconducting at $x \approx 0.20$. As x is increased to optimum, the material, originally an insulator, converts to a doped in-plane 2D metal with strong ionic features along the c-axis. Further doping results in the material transforming into a normal 3D metal. It is conceivable that multiphonon processes are enhanced by the transition to a normal metal.

The sticking between theory and experiment is further tested by deriving the scaling P factors for both $\rho_c$ and $\rho_{ab}$, as in Table I. We apply the theoretical formula to an uniform distribution of Cu(P)-O(A) bonds: $P_{theor}=k_B/ze\,N'xpR_{Cu-O}$ where $z$ is the coordination number at the Cu site (numer of Cu-O bonds). The doped hole density is $N = 10^{22}n_{orb}x \equiv N'x$ where $n_{orb}$ is the relative population of the oxygen orbital: $n_{x^2-y^2}$ in-plane and $n_{z^2}$ axial. Setting $p = eR_{Cu-O}$, we get $P_{theor}=4.68\times 10^7/x$ [$\Omega$cm/Ks] for $\rho_c$ ($z=2$, $R_{Cu-O}=2.4$E, $n_{z^2}=0.1$). Extending the model, we get $P_{theor}= 4.16\times 10^7/x$ [$\Omega$cm/Ks] for $\rho_{ab}$ ($z=4$, $R_{Cu-O}=1.8$E, $n_{x^2-y^2} =1$). $P_{fit}$ was calculated by linear regression in $T/\rho$ vs. $k_{12}(T)_{theor}$ co-ordinates with $k_{12}(T)_{theor}$ computed using the reaction-rate formula at fitting parameters $\varepsilon_B$, $\eta$, and $\hbar\omega$. It should be noted that for non-adiabatic (small $\eta$) narrow-gap electron transfers the electronic probabilities at both underbarrier and overbarrier levels are simply $W_{el}(n) = 4\pi\gamma_n$ (at $\gamma_n \ll 1$ viz. $\eta \ll 1$). We remind that $\gamma_n \propto \eta^2$ so that $\eta$ too enters as a scaling factor to the non-adiabatic transfer rate $k_{12}(T)$. Concomitantly, the apparent fitting factor $P_{fit}$ should now compose of two scaling terms: (i) $P_{theor}$ incorporating both the conversion of mesuring units and doping effects and (ii) $\eta^2$; this gives $P_{fit} \sim P_{theor} \eta^2$. From the tabulated values, $P_{theor}$ appears underestimated in the $\rho_{ab}$ case. The misfit. is not surprising in view of the specific (mosaic) geometry of the $CuO_2$ planes where intersquare links are secured by shared O(P) elements. Alternatively, one should consider suggestions for a bipolaron nature of the in-plane transport. On the other hand, the better adherence to $P_{fit}$ in the $\rho_c$ case may be improved considering the lower occupancy of the $3d_z^2$ Cu orbital (up to 10% of the doped hole density).

5. General discussion

The vibronic Hamiltonian (4) is appropriate for describing phenomena in which a purely ionic picture suffices, such as the optical spectra. However, an ionic picture may become inadequate for dealing with transport phenomena dominated by hybridization of the relevant orbitals. Now replacing the basis eigenstates, e.g. $\{|d\rangle,|p\rangle\}$, by hybridized states, such as the bonding (B), nonbonding (NB) and antibonding (AB) states of the O(A)-Cu(P)-O(A) triatomic molecule may prove appropriate. If so, the ionic energy gap $\varepsilon_{dp}$ is to be replaced by one of the hybridized gaps $\varepsilon_{NB-AB,B-NB} = S\{\varepsilon_{dp} \pm \sqrt{(\varepsilon_{dp}^2 + 8t_{pd}^2)}\}$, where the lower (upper) sign refers to the separation between the B-NB and NB-AB states, respectively.[7,20] Alternatively, the gap may take another form if the electronic correlations are essential: $\varepsilon_{gap} = U - \Delta_{CT}$ where $U$ is the correlation energy and $\Delta_{CT}$ is the charge transfer gap. It implies that the hole transfer is between the Cu lower Hubbard band and the O band.

The energy gap $\varepsilon_{gap}$ is the fitting parameter depending directly on the energy spectrum of the excitations that drive the axial transport process. Accordingly, we shall next consider two models for $\varepsilon_{gap}$ based on: (i) the energy spectrum of the hybridized O(A)-Cu(P)-O(A) triatomic molecule and (ii) the energy spectrum incorporating the $e$-$e$ correlations splitting it into Hubbard bands.

We described an approach to the axial transport problem based on hole hopping between Cu(P) $3d_z^2$ and O(A) $2p_z$ orbitals. In view of the occupancy of the $3d_z^2$ orbital being only a few percent of the doped hole density, objections may be raised as to the feasibility of our calculations. However, it is easy to see that our model is more general and similar analyses can be made based on hole hopping between apex oxygen $a_{2u}$ orbitals and the planar $a_1$ or $b_1$ molecular orbitals, instead. The $2p_z$-$3d_z^2$ choice has the advantage of avoiding the Hubbard splitting associated with the inhibited double occupancy of the $3d_{x^2-y^2}$ level: The large positive U should not be that large when the two holes reside on different $3d_z^2$ and $3d_{x^2-y^2}$ orbitals.

Our fitting and derivative parameters at various $x$ are listed in Table I. We first compare data therein with cluster calculations which point to Jahn-Teller energies $\varepsilon_{JT}$ that fall within the range of tenths of an eV ($\varepsilon_{JT} > 0.6$ eV). However, the Table I data, as derived from the resistivities, suggest $\varepsilon_{JT}$ a hundred times lower (typically $\varepsilon_{JT} \sim 5$ meV). This implies that hybridization of the eigenstates plays an essential role in the carrier transport along the c-axis, as it should. Indeed, taking into account the electron hopping, e.g. $t_{b1-a2} \sim 0.015$ eV (cf. Fig.2 left), the bonding-to-nonbonding (B-NB) energy gap is $\varepsilon_{B-NB} = S|\Delta_{b1-a2} - \sqrt{(\Delta_{b1-a2})^2 + 8(t_{b1-a2})^2}|$ (~0.1meV), as for a triatomic O(A)-Cu(P)-O(A) molecule. This $\varepsilon_{B-NB}$ is the energy gap in the basis of hybridized symmetric and antisymmetric combinations of the $b_1$ and $a_2$ molecular orbitals of the $La_8CuO_6$ cluster. Plausibly, $\varepsilon_{B-NB}$ should enter as an energy gap $\varepsilon_{gap}$ in the vibronic mixing of the hybridized levels by $A_{2u}$ and $E_u$ modes to generate double-well potentials and thereby the dynamic coupling which builds the temperature dependence of the resistivity. From a reaction-rate viewpoint, the low gap energies $\varepsilon_{gap} \sim 0.8$ meV leading to $\eta \sim 0.03$ imply a nonadiabatic electron transfer.

A first glance point arising from the fitting parameters in Table I is the coupled axial phonon frequency which appears associated with an acoustic rather than optical vibration. We have no comment at this point of the acoustic character, though we remind that the low fitting frequency comes from the low energy meV scale of our transport parameters.

We also see that the interwell barriers $\varepsilon_B$ do not differ drastically on going from $\rho_c$ to $\rho_{ab}$. Nevertheless, the barriers seem to first jump and then drop, as the doping $x$ is increased towards the overdoped range where the resistivity turns purely metallic. A mostly metallic resistivity would cover the entire temperature range near $T$ for barriers so low that $\varepsilon_B \ll k_B T$. The crossover (polaron binding) energy $\varepsilon_C \approx \varepsilon_B$ being inversely proportional to the polaron size, an ultimate axial conversion is implied from smaller polarons to larger polarons as the carrier concentration $x$ goes to the overdoped range. This expectation is convergent with our earlier two-fluid boson-fermion model for the $T_c$-$p$ phase diagrams of $La_{2-x}Sr_xCuO_4$.[20] A similar trend of increasing the polaron size may be deduced from the in-plane data.

Earlier literature parameter data are summarized in Table II.[7,20,28,29] For comparison, we present a summary of vibronic-polaron parameters reported in preceding papers, as derived from optical and transport data on $La_{1-x}Sr_xCuO_4$. These parameters are tabulated along with our assignments.

The temperature dependences by Boebinger *et al.* are remarkable in several ways.[13] First, the experimental $\rho_c$ and $\rho_{ab}$ data therein show that the dominant feature below or near
optimum doping in the superconducting range ($T < T_c$), where pairing has been suppressed by an external magnetic field, is the scattering of fermionic excitations by double wells. We believe this conclusion is an essential clue to the superconductive mechanism. Second, experimental $\rho_c$ data show that the same excitations active in the magnetically-suppressed-pairs range below $T_c$ are those operative in normal state above $T_c$. Otherwise there would have been a change of thermal slope across $T_c$, not displayed by the $\rho_c$ data. This lends support to a BCS-like pairing mechanism of vibronic polarons along the c-axis, which polarons form as fermionic excitations couple to the double wells, rather than pairing as real-space vibronic bipolarons, since BCS pairs occur and condense at $T_c$, while bipolarons should be there at $T > T_c$. The vibronic polarons evolve from small through large, as doping is increased from the underdoped range across the optimum doping through the overdoped range.[20,28] Third, the experimental $\rho_{ab}$ data do not lead to similar conclusions, for there is an apparent change of slope and magnetoresistance in the crossover range. The $\rho_{ab}$ data do not rule out that the in-plane transport in normal state is dominated by a double fluid of polarons and bipolarons, as in double-layered cuprates. In normal state at $T > T_c$, the behavior near optimum doping and above is rather metallic both in-plane and out-of-plane.

The present microscopic study of double-well (two-level) systems is in concert with path-integral investigations of similar species.[8] The double- well results suggest that a coherent axial transport may arise from the coupling of fermionic excitations to site-splitting potentials associated with apex oxygens. Our model predictions agree with the experimental temperature dependences of the axial resistivity.[4,13] Yet, actual phase coherence may be lost by virtue of an intense carrier scatter preceding the otherwise coherent axial hops.[9] It should also be stressed that the present c-axis transport in the strong-coupling limit is presumably carried by non-degenerate polarons. The average kinetic energy of a non-degenerate electronic gas is $k_B T$, while it is $\varepsilon_F$ in the degenerate electronic gas of a metal. This explains why the Fermi energy $\varepsilon_F$ is missing in our *c*-axis equations, unlike the situation in *ab*-plane or in metals.

Finally, we stress that the present $\rho \propto T$ part of the resistivity dominant in the metallic range arises from our applying Boltzmann statistics to the averaged phonon-coupled traversal time $\tau_{12} = k_{12}^{-1}$ along the Cu-O bond. Boltzmann statistics had led Sir Nevill to deducing a related temperature dependence for the carrier mobility μσ thereby conductivity σσ from a temperature-independent diffusion coefficient Dσ of *spin polarons* using the Nernst-Einstein relation.[30] From μl in Section 2, our *dielectric bond-polarons* have a diffusion coefficient

$$D_l = (k_B T/e)\mu_l \sim (p/e)R_{Cu(P)-O(A)}k_{12}(T) \sim (R_{Cu(P)-O(A)})^2 k_{12}(\infty) \qquad (12)$$

that is also temperature-independent in the metallic range. It might be that the observed $\rho \propto T$ dependence implies that axial, possibly in-plane excitations too, migrate quasifree as bond polarons which do not undergo any other significant scattering or multiple trapping in normal state. The semiconductor-like resistivity range, as uncurtained in pulsed magnetic fields, provides an otherwise concealed evidence for the formation of these quasifree carriers.

In conclusion, we applied a model where the dominant feature shaping the temperature behavior of both the axial and in-plane electrical resistivities in single-layered high-temperature superconducting cuprates is the scattering of fermionic excitations by double-well potentials. The model appears not only to agree with qualitative features of the temperature dependences but also to stick quantitatively to the experimental data.

TABLE I. Present double-well and vibronic-polaron parameters for $La_{2-x}Sr_xCuO_4$
(all energies in meV)

| | In-plane | | | | | Axial | | | |
|---|---|---|---|---|---|---|---|---|---|
| $x$ | 0.08 | 0.12 | 0.15 | 0.17 | 0.22 | 0.12 | 0.13 | 0.0.15 | 0.17 0.22 |
| $\varepsilon_B$ | 4.74 | 5.61 | 2.90 | 3.39 | 2.45 | 6.94 | 5.51 | 9.75 | 7.36 5.11 |
| $\hbar\omega$ | 0.70 | 0.70 | 0.80 | 0.70 | 0,70 | 0.70 | 0.70 | 0.70 | 0.70 0.73 |
| $\eta$ | 0.042 | 0.033 | 0.049 | 0.043 | 0.072 | 0.029 | 0.035 | 0.022 | 0.031 0.027 |
| $\varepsilon_{JT}$ | 5.16 | 6.00 | 3.21 | 3.70 | 2.85 | 7.36 | 5.92 | 10.19 | 7.84 5.40 |
| $\varepsilon_{gap}$ | 0.86 | 0.79 | 0.63 | 0.63 | 0.83 | 0.85 | 0.83 | 0.89 | 0.97 0.59 |
| $\hbar\omega_{bare}$ | 0.73 | 0.72 | 0.84 | 0.73 | 0.75 | 0.72 | 0.73 | 0.72 | 0.72 0.75 |
| $\varepsilon_R$ | 20.60 | 23.97 | 12.81 | 14.77 | 11.34 | 29.42 | 23.65 | 40.74 | 31.3 21.58 |
| $\varepsilon_C$ | 5.17 | 6.01 | 3.22 | 3.71 | 2.86 | 7.37 | 5.93 | 10.19 | 7.85 5.40 |
| $K_{12}s^{-7}$ | 1.4 | 0.55 | 1.8 | 1.3 | 2.3 | 0.36 | 1.2 | 0.18 | 0.39 0.77 |
| $A^{-8}K^{-1}$ | 16 | 7.8 | 26 | 18 | 30 | 4.9 | 11 | 2 | 4.9 8.5 |

6. Pairiing of bond polarons and high-$T_c$ superconductivity in single layered $La_{2-x}Sr_xCuO_4$.

In the quest for a reliable, possibly true pairing mechanism for the high-$T_c$ oxocuprates, various models have been considered. This endeavor has been fed by the expectation that no conventional (electron-phonon) binding for these superconductors has been adequate in so far as it predicts transition temperatures of up to 10-15 K which is inferior to the measured ones. Ideologically closest to the material properties has been the superconductivity in Jahn-Teller (JT) materials. Then the exciton-mediated pairing has then been called forth to model unconventional processes, if anything. Soon the excitons have been ruled out too and the attention focused onto unconventional materials [24-32].

Binding in JT materials has been called forth first, soon after discovering a high-temperature superconductivity in experiments on $La_{2-x}Ba_xCuO_4$ (~30K) [1]. This has been the most natural explanation in view of the unusual electron-phonon character, and indeed, as the conventional phononic mechanism failed to produce an unconventional behavior, then why shouldn't we look for a phononic mechanism somewhere nearby, possibly in the phonon-mixing of electron states, or vibronic mixing as we often call it? Indeed, the phonon-mixing has been suspected in as much as JT exhibiting materials provided the best examples, such as

YBa$_2$Cu$_3$O$_7$ (~90K). Well conceived though not fully agreed with experiment. Then, in looking for other though neighboring ways, attention was drawn by Pseudo-JT (PJT) materials, species related though not completely to JT ones [2]. The high-Tc materials being non-conductive insulators at $x \sim 0$, become conductors at finite though moderate x on incorporating divalent ions at trivalent ion sites. See Figure 1 for the phase diagram of La$_{2-x}$Ba$_x$CuO$_4$ [3].

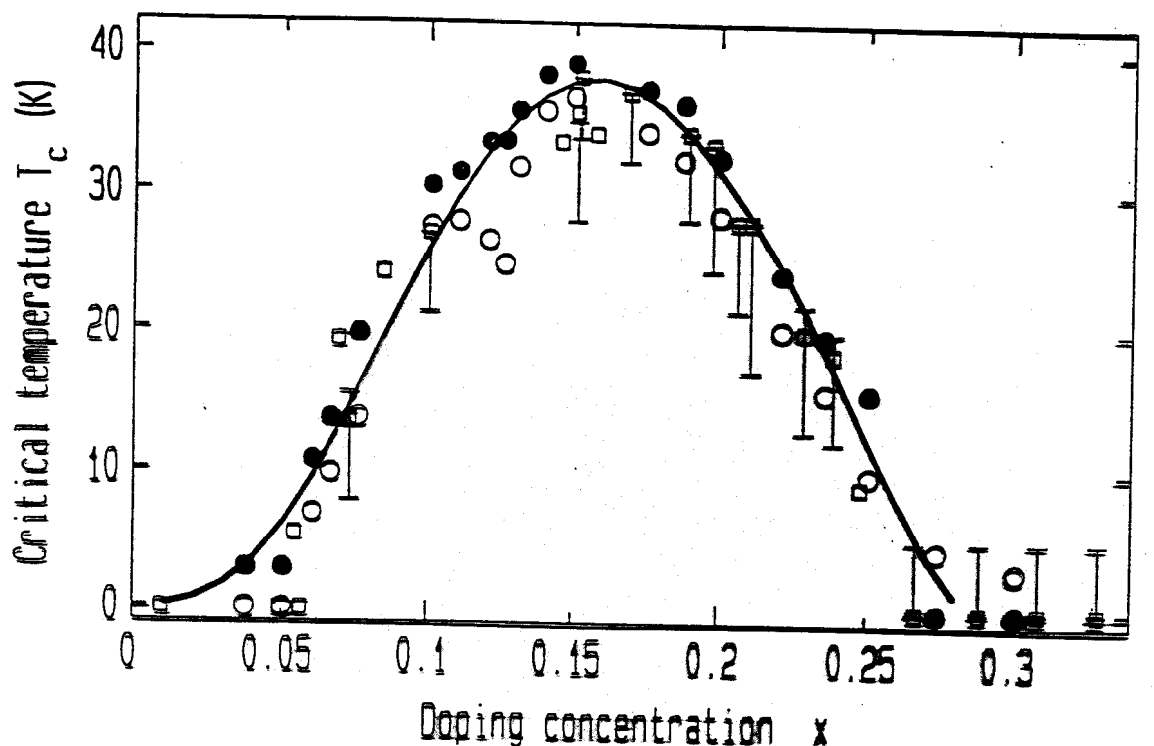

Figure 1: Showing the phase diagram of La$_{2-x}$Ba$_x$CuO$_4$ (Ref. [3,20]).

The diagram plots the critical temperature of superconductivity vs the doping concentration $x$. It shows a maximum $T_C \sim 38$ K at a doping concentration of 0.15. Unlike JT units, which do not exhibit any electrostatic dipole of their own, PJT units do. This is the inversion dipole occurring as a result of breaking the inversion symmetry at a polaron site. Polarons form as a charge carrier (hole) becomes bound to a CuO6 octahedron or to to a Cu-O bond at an octahedron. It is these polarons (basic charge carriers) which give rise to the peculiar properties of these high-Tc materials. We call them *polarons* although bond and free polarons may differ substantially from each other even though coexisting in a real material.

Fugure 2: Showing a selected example of the temperature dependent rate constant of in-plane currents in LaSCO [32].

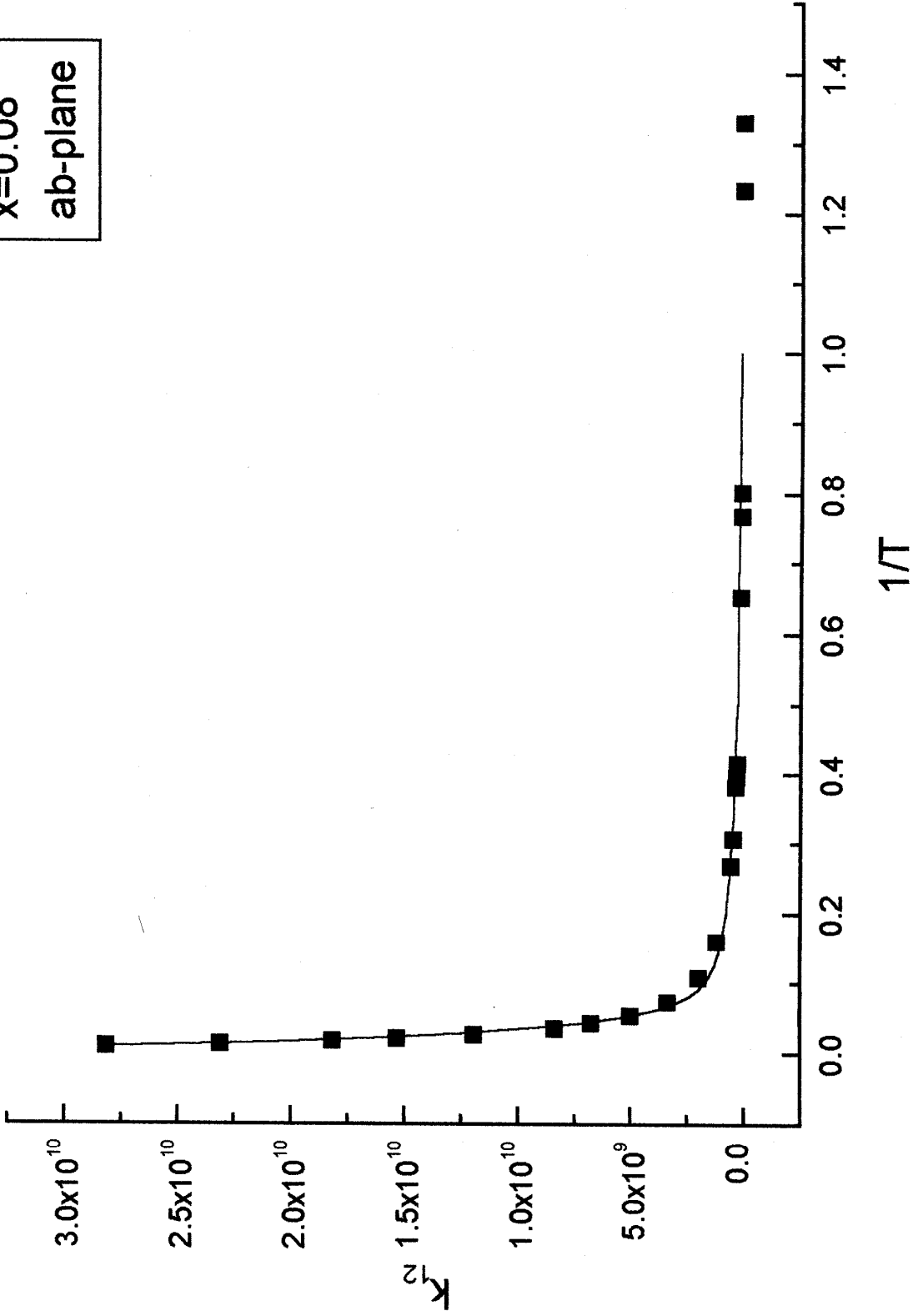

As far as pairing is concerned, the occurrence of inversion dipoles at the polaron site makes applying the formalism of dispersive interactions and dispersive forces directly applicable. We postpone for a later discussion the detail considerations and general theory of VdW forces based on *Fritz London's* model. Here we only outline the most general features of the model as well as its direct relationship to the bond polarons and their experimental behavior.

6.1. Inversion dipoles

A basic building block of high-temperature superconductors of the $La_{2-x}Ba_xCuO_4$ family is the $CuO_6$ octahedron. We conceive various symmetries of the $CuO_6$ unit both symmetric (cubic $O_h$ symmetry) and asymmetric (e.g. $A_{2u}$ or $E_u$) forms. The former of the asymmetric forms gives rise to an inversion dipole. It forms along the z- (c-) axis in two modifications, up- and down- Cu-O dipoles. Attractive binding is feasible between up- and down- forms only, combinations like up-up and down-down leading to dipolar repulsion. For a simple estimate of a dipole-dipole coupling of species arising in neighboring octahedrons we set $\alpha = p_0 \cdot p_0 / \Delta$ for the dipolar polarizability and $\Delta \alpha^2 / \kappa R^3$ for the coupling energy where $\Delta$ is the interlevel energy gap, R is the inter- Cu separation. In as much as essentially the interaction is between two neighboring dipoles (not necessarily nearest neighbors), the coupling is of two inversion dipoles and hence it may be considered to be an estimate of the inversion-dipole pairing energy. Just how it would look like in the stereo-picture of the $CuO_6$ octahedrons, we refer to the literature. We see the configurational picture of the coupling between two dipoles but if fact this coupling may give rise to London's dispersive force within a dipolar field between polarizable though neutral species whereby the inversion dipoles are fluctuating to give rise to dispersive force like the one between s-atoms in atomic physics.

Tables exhibiting fitting parameters obtained as barrier controlled normal state currents have been worked out by means of the axial phonon coupled rates. as shown in Figure 3. One of these parameters is the barrier energy which is estimated on the average at ~ 7 meV for the axial currents and at ~ 4 meV for the in-plane currents. Nearly twice as much barrier energy has been found necessary for the axial process to materialize as compared to the in-plane process. We interpret the barrier energy to be a measure for the pairing energy which may be a good guess in view of the above numerical estimates for the latter. Indeed with an average gap energy of 0.8 meV for $\Delta$ and $\alpha \sim R^3$, we obtain $\varepsilon_B = \Delta \alpha^2 / \kappa^2 R^6 \sim 40$ meV close to fitting.

The foregoing estimate is based on applying the Van-der-Waals (VdW) coupling energy which seems to be most appropriate in view of the dispersive-like character of the interaction. Nevertheless, London's model for the pairing energy is based on a $2^{nd}$ order perturbation analysis and may not be appropriate for polarizabilities largely exceeding the unit cell volume

No boundary has yet been defined which poses a problem to occupy theoreticians for a time.

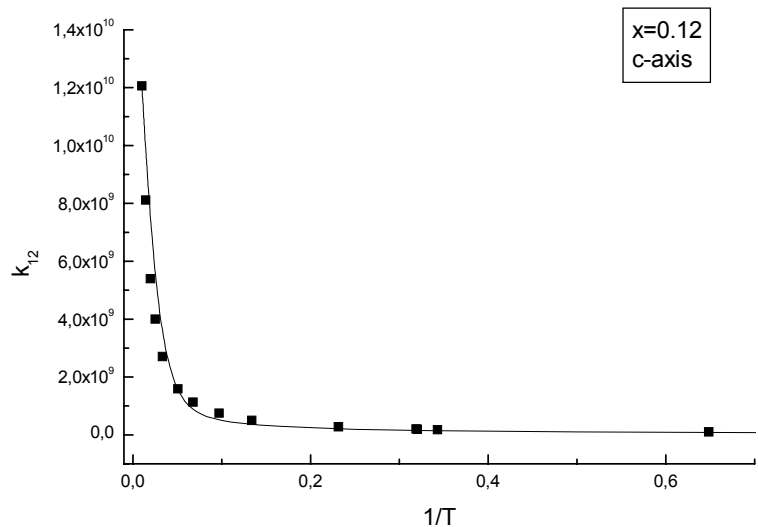

Figure 3: Temperature dependence of the axial rate constant in LaSCO conductor [32]